\documentclass[twocolumn,aps,superscriptaddress,showpacs,floatfix]{revtex4}
\usepackage{graphicx}
\usepackage{dcolumn}
\usepackage{bm}
\usepackage[colorlinks,linkcolor=blue,anchorcolor=blue,citecolor=blue,urlcolor=blue]{hyperref}
\usepackage{amsmath}
\usepackage{multirow}
\usepackage{booktabs}  
\usepackage{longtable}  
\usepackage{float}
\usepackage{array}
\usepackage{etoolbox}
\usepackage{makecell,rotating,multirow,diagbox}
\usepackage{setspace}
\usepackage{tabularx}
\usepackage[draft]{changes}
\usepackage[section]{placeins}
\usepackage{lineno,hyperref}
\usepackage{supertabular}
\usepackage{amssymb}
\usepackage[normalem]{ulem}
\usepackage[colorlinks,linkcolor=blue,anchorcolor=blue,citecolor=blue,urlcolor=blue]{hyperref}
\usepackage[thicklines]{cancel}
\usepackage{xcolor}

\renewcommand\sout{\bgroup \color{red} \ULdepth=-.5ex \ULset}

\begin{document}

\title{An analytic formula for the proton radioactivity spectroscopic factor}

\author{Dong-Meng Zhang}
\affiliation{School of Nuclear Science and Technology, University of South China, 421001 Hengyang, People's Republic of China}
\author{Lin-Jing Qi}
\affiliation{School of Nuclear Science and Technology, University of South China, 421001 Hengyang, People's Republic of China}
\author{Hai-Feng Gui}
\affiliation{School of Nuclear Science and Technology, University of South China, 421001 Hengyang, People's Republic of China}
\author{Song Luo}
\affiliation{School of Nuclear Science and Technology, University of South China, 421001 Hengyang, People's Republic of China}

\author{Biao He}
\email{hbinfor@163.com}
\affiliation{College of Physics and Electronics, Central South University, Changsha 410083, China}

\author{Xi-Jun Wu}
\email{wuxijun1980@yahoo.cn}
\affiliation{School of Math and Physics, University of South China, Hengyang 421001, People's Republic of China}

\author{Xiao-Hua Li}
\email{lixiaohuaphysics@126.com }
\affiliation{School of Nuclear Science and Technology, University of South China, 421001 Hengyang, People's Republic of China}
\affiliation{Cooperative Innovation Center for Nuclear Fuel Cycle Technology $\&$ Equipment, University of South China, 421001 Hengyang, People's Republic of China}
\affiliation{National Exemplary Base for International Sci $\&$ Tech. Collaboration of Nuclear Energy and Nuclear Safety, University of South China,Hengyang 421001, People's Republic of China}
\begin{abstract}
	
In the present work, we systematically study the spectroscopic factor of proton radioactivity ($S_p$) with $A>100$ using the deformed two-potential approach (D-TPA). It is found that there is a link between the quadrupole deformation parameter of proton emitter and $S_p$. Based on this result, we propose a simple analytic formula for estimating the spectroscopic factor of proton radioactivity. With the help of this formula, the calculated half-lives of proton radioactivity can reproduce the experimental data successfully within a factor of 2.77. Furthermore, we extend the D-TPA with this formula for evaluating the spectroscopic factor to predict the proton radioactivity half-lives of 12 proton radioactivity candidates whose radioactivity is energetically allowed or observed but not yet quantified in NUBASE2020. For comparison, the universal decay law for proton radioactivity (UDLP) and the new Geiger-Nuttall law (NG-N) are also used. It turns out that all of the predicted results are basically consistent with each other.

\end{abstract}

  \maketitle
\section{Introduction}
\setlength{\parskip}{0pt}
One of the most important decay modes for proton-rich nuclide far away from $\beta$-stability line is proton radioactivity, which is a quantum-tunneling effect \cite{C. Xu Phys. Rev. C 2016}. Studying proton radioactivity half-lives may be of great importance in promoting the existing nuclear theories and models, describing nuclear deformation, and understanding the decay properties of nuclei \cite{Delion D S Phy. Rev. Lett. 2006,Blank B Prog. Part. Nucl. Phys. 2008,Zhang H F J. Phys. G: Nucl. Part. Phys. 2010,Chen J L J. Phys. G: Nucl. Part. Phys. 2019,Delion D S Phys. Rev. C 2009,Karny M Phys. Lett. B 2008,Zhang Z X Chin. Phys. C 2018,Basu D N Phys. Rev. C 2005}. The first confirmation for proton radioactivity was obtained by Jackson \emph{et al.} in the measurement of an isomeric state for $^{53}$Co \cite{Jackson K Phys. Lett. B 1970,Cerny J Phys. Lett. B 1970}. In 1981, at the velocity filter SHIP at GSI, Hofmann \emph{et al.} first discovered the proton emission from nuclear ground state of $^{151}$Lu \cite{Hofmann S Z. Phys. A 1982}. Subsequently, the proton emitter $^{147}$Tm was discovered in experiment performed at GSI \cite{O. Klepper Z. Phys. A: At. Nucl. 1982} as well as $^{109}$I and $^{113}$Cs were observed using a catcher foil technique at Munich \cite{T. Faestermann Phys. Lett. B 1984}. In 1984, Hofmann \emph{et al.} discovered the isomeric decay in $^{147}$Tm and the ground decay of $^{150}$Lu \cite{S. Hofmann 1984}. The highly deformed proton emitters $^{141}$Ho and $^{131}$Eu were reported by Davids \emph{et al.} in subsequent years \cite{C. N. Davids Phys. Rev. Lett. 1998}. In addition to the above decay processes, with the advancement of diverse infrastructures and radioactive beam installations, a number of proton emitters have been provenly illustrated in the proton regions $Z=51-$83  from the ground state or low-lying isomeric state during the last decades \cite{Santhosh K P Pramana J. Phys.,Zhang H F Sci. China. Ser. G:Phy. Mech. Astron.}. 

Up until now proton radioactivity has become one of the most popular topics in the field of nuclear physics \cite{Delion D S Phy. Rev. Lett. 2006,Blank B Prog. Part. Nucl. Phys. 2008,Delion D S Phys. Rev. C 2009,Y. Z. Wang Commun. Theor. Phys. 2021}. Moreover, study on the spectroscopic factor of proton radioactivity may facilitate the investigation of nuclear structures such as nuclear deformation \cite{Qian Y and Ren Eur. Phys. J. A 2016,L.S. Ferreira PhysRevC.65.024323} and quantum mixing \cite{P. J. Woods 1997}. In theory, the spectroscopic factor of proton radioactivity($S_p$) can be regarded as a probability that the blocked proton in the orbit of the parent nucleus would transfer to the empty orbit of the daughter nucleus. There are many microscopical theories to estimate it\added{,} including the relativistic mean field theory (RMF) with the Bardeen-Cooper-Schriffer (BCS) model \cite{Zhang H F J. Phys. G: Nucl. Part. Phys. 2010,Qian Y and Ren Eur. Phys. J. A 2016,J. M. Dong Phys. Rev. C 2009,D. Delion Phys. Rep. 2006,A.Soylu Chin. Phys. C 2021}, relativistic continuum Hartree-Bogoliubov (RCHB) theory \cite{Y. Lim Phys. Rev. C 2016}, covariant density functional (CDF) with BCS method \cite{Q. Zhao Phys. Rev. C 2014}, etc. Phenomenologically, $S_p$ can be obtained by the ratio of the theoretical proton radioactivity half-life to experimental data \cite{Chen J L et al Eur. Phys. J. A 2021}. Besides, in the numerous calculations, $S_p$ is generally assumed to be a constant \cite{Chen J L J. Phys. G: Nucl. Part. Phys. 2019,Santhosh K P Pramana J. Phys.}, which causes the microscopic information about the nuclear structure not to be reflected. 

Recently, Chen \emph{et al.} \cite{Chen J L et al Eur. Phys. J. A 2021} proposed a simple analytic formula for the spectroscopic factor of spherical proton emitters with the same orbital angular momentum $l$. However, this formula is unsuitable for the well deformed proton emitters. In 2021, Delion \emph{et al.} discovered that there is a small residual linear decrease of the logarithm of the spectroscopic factor versus the quadrupole deformation parameter of proton emitters \cite{D. S. Delion Phys. Rev. C 2021}. As a result, it is an interesting topic from a new perspective to propose a unified definition of the spectroscopic factor for spherical and deformed proton emitters. To account for this, we establish a connection between the spectroscopic factor and the deformation parameter employing the deformed two-potential approach (D-TPA). The results suggest that $S_p$ can be expressed as a formula involving the quadrupole deformation parameter of proton emitters. This formula can be utilized for estimating the spectroscopic factor and conducting accurate calculations of proton radioactivity half-lives.

This article is organized as follows. In Section \ref{section 2}, the theoretical framework of D-TPA is introduced in detail. The results and discussion are presented in Section \ref{section 3}. Finally, a brief summary is given in Section \ref{section 4}. 

\section{Theoretical framework}
\label{section 2}
The proton radioactivity half-life $T_{1/2}$ is correlated with the decay width $\Gamma$. It can be expressed as 
\begin{eqnarray}
T_{1/2}=\frac{\hbar \ln2}{\rm{\Gamma}},
\label{eq1}
\end{eqnarray}
where $\hbar$ represents the reduced Planck constant. In the framework of two potential approach (TPA) \cite{Chen J L et al Eur. Phys. J. A 2021,J. H Cheng Phys. Rev. C 2022,H. F. Gui Commun. Theor. Phys. 2022}, $\rm{\Gamma}$ can be determined by the normalized factor $F$ and penetration probability $P$, and expressed as
\begin{eqnarray}
{\rm{\Gamma}}=S_p\frac{\hbar^2FP}{4\mu},
\label{eq2}
\end{eqnarray}
where $S_p$ is the spectroscopic factor of the emitted proton-daughter system. In the present work, $S_p$ is assumed as a constant, i.e., $S_p=1$, which is crucial for the subsequent systematic analysis. $\mu=mA_d/(A_d+A_p)\ {\rm MeV/c^2}$ is the reduced mass of the emitted proton-daughter nucleus system, where $m$ is the nucleon mass. $A_p$ and $A_d$ are the mass number of the emitted proton and daughter nucleus, respectively. 

$F$ represents the normalized factor, which describes  the emitted proton assault frequency as it passes through the potential barrier. Taking into account the effect of deformation, we can obtain the total normalized factor $F$ by averaging $F_{\theta}$ over all possible orientations. It can be expressed as
\begin{eqnarray}
F=\frac{1}{2}\int_0^{\pi}F_{\theta}\sin\theta{d\theta},
\label{eq3}
\end{eqnarray}
where $\theta$ is the orientation angle of the emitted proton with measured from the symmetry axis of the daughter nucleus and $F_{\theta}$ can be written as
\begin{eqnarray}
F_{\theta}=\frac{1}{\int_{r_1}^{r_2}\frac{1}{2k(r,\theta)}{dr}}.
\label{eq4}
\end{eqnarray}

$P$, the penetration probability for the emitted proton penetrating the barrier, can be obtained by the semiclassical WKB approximation. Considering the effect of deformation, we can obtain the total penetration probability $P$ by averaging $P_{\theta}$ in all direction. It can be written as 
\begin{eqnarray}
P=\frac{1}{2}\int_0^{\pi}P_{\theta}\sin\theta{d\theta},
\label{eq5}
\end{eqnarray}
where the polar-angle-dependent penetration probability of proton radioactivity $P_{\theta}$ is given by
\begin{eqnarray}
P_{\theta}={\rm{exp}}\left[-2\int_{r_2}^{r_3}k(r,\theta){dr}\right].
\label{eq6}
\end{eqnarray}
Here $k(r,\theta)=\sqrt{\frac{2\mu}{\hbar^2} \vert V(r,\theta)- Q_p \vert }$ is the wave number. $V(r, \theta)$ is the total interaction potential between the emitted proton and daughter nucleus. In Eq. \ref{eq4} and \ref{eq6}, $r_1$, $r_2$ and $r_3$ represent the classical turning points, which satisfy the condition $V(r,\theta)=Q_p$. The proton radioactivity released energy $Q_p$ can be obtained by 
\begin{eqnarray}
Q_p=\Delta M-(\Delta M_d+\Delta M_{p})+k(Z^{\varepsilon}-Z_d^{\varepsilon}),
\label{eq7}
\end{eqnarray}
where $\Delta M, \Delta M_d$ and $\Delta M_p$ represent the mass excesses of parent, daughter nucleus and the emitted proton, respectively. They are taken from the latest atomic mass table NUBASE2020 \cite{Kondev F G Chin. Phys. C 2021}.
The term $k(Z^{\varepsilon}-Z_d^{\varepsilon})$ denotes the screening effect of atomic electrons with $k=13.6$ eV, $\varepsilon=2.408$ for $Z<60$, and $k=8.7$ eV, $\varepsilon=2.517$ for $Z\geq 60$. Here $Z$ and $Z_d$ are the proton number of parent nucleus and daughter nucleus, respectively. \cite{Denisov V Y Phys. Rev. C 2005,K. N. Huang 1976}.

In the present work, the total interaction potential $V(r,\theta)$ consists of the nuclear potential $V_{N}(r,\theta)$, Coulomb potential $V_{C}(r,\theta)$ and centrifugal potential $V_{l}(r)$. It can be expressed as
\begin{eqnarray}
V(r,\theta) = V_{N}(r,\theta) + V_{C}(r,\theta) + V_{l}(r).
\label{eq8} 
\end{eqnarray}
The emitted proton-daughter nucleus nuclear potential $V_{N}(r,\theta)$ is chosen as a type of cosh parametrized form \cite{B. Buck Phys. Rev. C 1992}. It can be expressed as
\begin{eqnarray}
V_N(r,\theta)=-V_0\frac{1+\cosh(R_d(\theta)/a)}{\cosh(r/a)+\cosh(R_d(\theta)/a)},
\label{eq9}
\end{eqnarray}
where $V_0$ and $a$ are the parameters of the depth and diffuseness of the nuclear potential, respectively. We choose $V_0=57.83$ MeV and $a=0.857$ fm, which are taken from Ref.\cite{J. H Cheng Phys. Rev. C 2022}. Taking into account the deformed effect of daughter nucleus, the $R_d(\theta)$ is given by \cite{M. Ismail  Nucl. Phys. A 2017}
\begin{eqnarray}
R_d(\theta)=R_d^{'}(1+\beta_2Y_{20}(\theta)+\beta_4Y_{40}(\theta)+\beta_6Y_{60}(\theta)),
\label{eq10}
\end{eqnarray}
where $\beta_2$, $\beta_4$ and $\beta_6$ denote the quadrupole, hexadecapole and hexacontatetrapole deformation parameters of the residual daughter nucleus, which are taken from FRDM2012 \cite{A. J. Sierk 2016}. $Y_{lm}(\theta)$ is a spherical harmonics function. $R_d^{'}$ is the spherical radius of daughter nucleus \cite{N. Wang and W. Scheid Phys. Rev. C 2008}, which can be expressed as
\begin{eqnarray}
R_d^{'}=1.27A_d^{1/3}.
\label{eq11}
\end{eqnarray}

Based on the double folding model, the Coulomb potential for the deformed daughter and emitted proton can be expressed as \cite{N. Takigawa Phys. Rev. C 2000}
\begin{equation}
V_{C}(\vec{r},\theta)=\iint\frac{\rho_d(\vec{r_1})\rho_p(\vec{r_2})}{\vert\vec{r}+\vec{r_1}+\vec{r_2}\vert}d\vec{r_1}d\vec{r_2},
\label{eq12} 
\end{equation}
where $\vec{r}$ is the vector between the centers of the emitter proton and daughter nucleus. Meanwhile, $\vec{r_1}$ and $\vec{r_2}$ represent the radius vectors in the charge distributions of the emitter proton and daughter nucleus, respectively. $\rho_d(\vec{r_1})$ and $\rho_p(\vec{r_2})$ denote the charge density distribution of the deformed daughter nucleus and spherical emitted proton, respectively. Simplified by the Fourier transform, the Coulomb potential can be presented as \cite{N. Takigawa Phys. Rev. C 2000,M. Ismail Phys. Lett. B 2003,G. L. Zhang Chin. Phys. Lett. 2008}
\begin{equation}
V_{C}(\vec{r},\theta)=V_{C}^{(0)}(\vec{r},\theta)+V_{C}^{(1)}(\vec{r},\theta)+V_{C}^{(2)}(\vec{r},\theta),
\label{eq13} 
\end{equation}
where $V_{C}^{(0)}(\vec{r},\theta)$, $V_{C}^{(1)}(\vec{r},\theta)$ and $V_{C}^{(2)}(\vec{r},\theta)$ are the bare Coulomb
interaction, linear Coulomb coupling, and second-order Coulomb coupling, respectively. The specific expressions of $V_{C}^{(0)}(\vec{r},\theta)$, $V_{C}^{(1)}(\vec{r},\theta)$ and $V_{C}^{(2)}(\vec{r},\theta)$ are presented in Ref. \cite{H. F. Gui Commun. Theor. Phys. 2022}.

For the centrifugal potential $V_l(r)$, we adopt the Langer modified form, since the correction $l(l+1)\rightarrow (l+1/2)^2$ is essential for one-dimension problems \cite{Morehead J J J. Math. Phys. 1995}. It can be expressed as
\begin{eqnarray}
V_{l}(r)=\frac{\hbar^2(l+\frac{1}{2})^2}{2\mu r^2}.
\label{eq14} 
\end{eqnarray}
Here $l$ is the angular momentum taken away by the emitted proton, which satisfies the spin-parity conservation laws. It can be expressed as
\begin{eqnarray}
l=
\left\{
\begin{array}{ll}
\Delta_j & \quad \mbox{for even}\ \Delta_j \ \mbox{and}\ \pi=\pi_d,\vspace{0.3em}\\
\Delta_j+1 & \quad \mbox{for even}\ \Delta_j \ \mbox{and}\ \pi\neq\pi_d,\vspace{0.3em} \\  
\Delta_j & \quad \mbox{for odd}\ \Delta_j \ \mbox{and}\ \pi\neq\pi_d,\vspace{0.3em} \\
\Delta_j+1 & \quad \mbox{for odd}\ \Delta_j \ \mbox{and}\ \pi=\pi_d,
\end{array}
\right.
\label{eq15} 
\end{eqnarray}
where $\Delta_j=\vert j-j_d-j_p \vert$ with $j$, $\pi$, $j_d$, $\pi_d$, and $j_p$, $\pi_p$ denoting the spin and parity values of the parent nucleus, daughter nucleus and the emitted proton, respectively.

\section{Results and discussion}
\label{section 3}

\begin{figure}[htb]\centering
	\includegraphics[width=8.7cm]{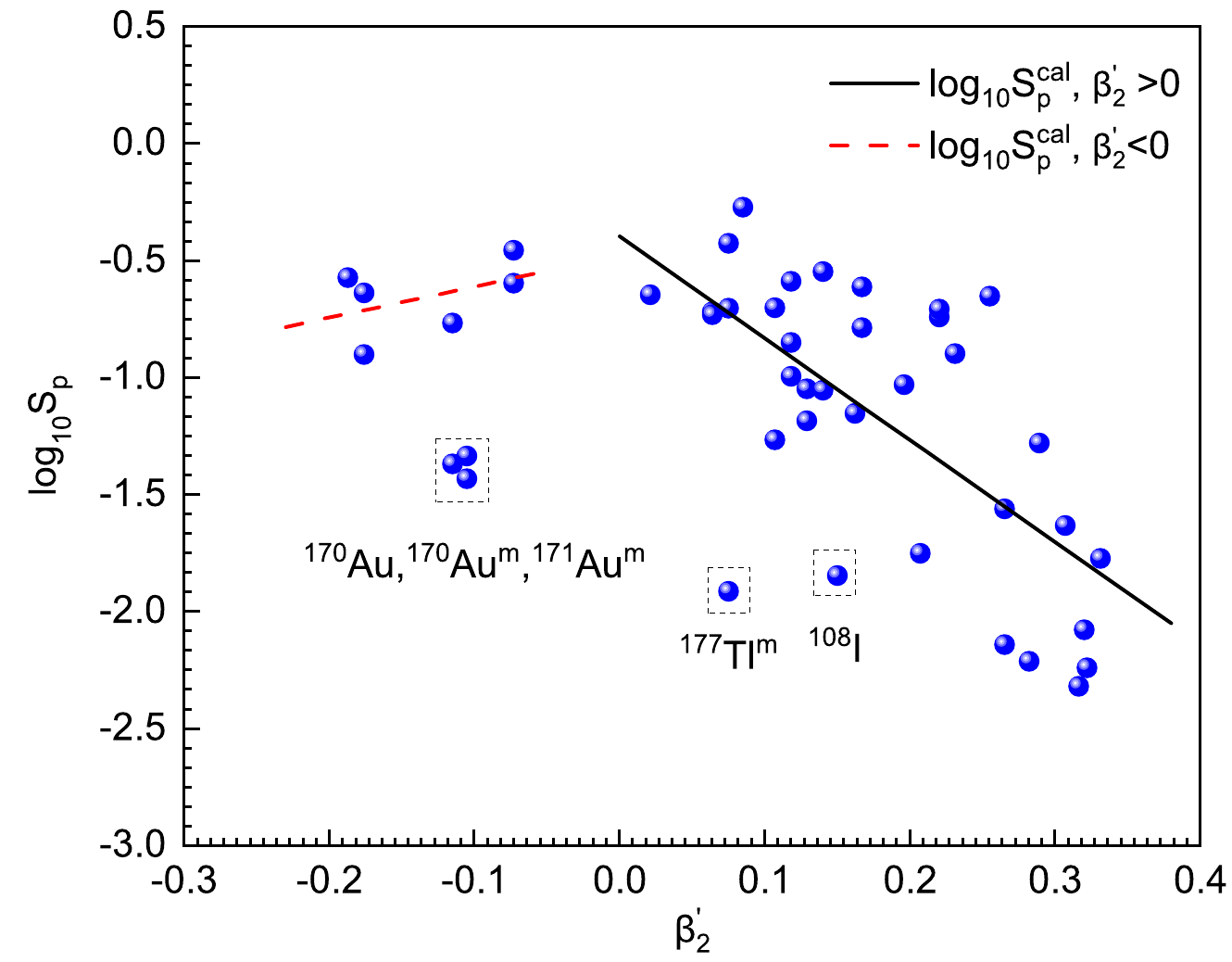}
	\caption{(color online) The linear relationship between the logarithmic value of experimental spectroscopic factor ${\rm log}_{10}S_p^{\rm exp}$ and $\beta_2^{'}$.}
	\label{fig 1}
\end{figure}

\begin{figure}[htb]\centering
	\includegraphics[width=8.5cm]{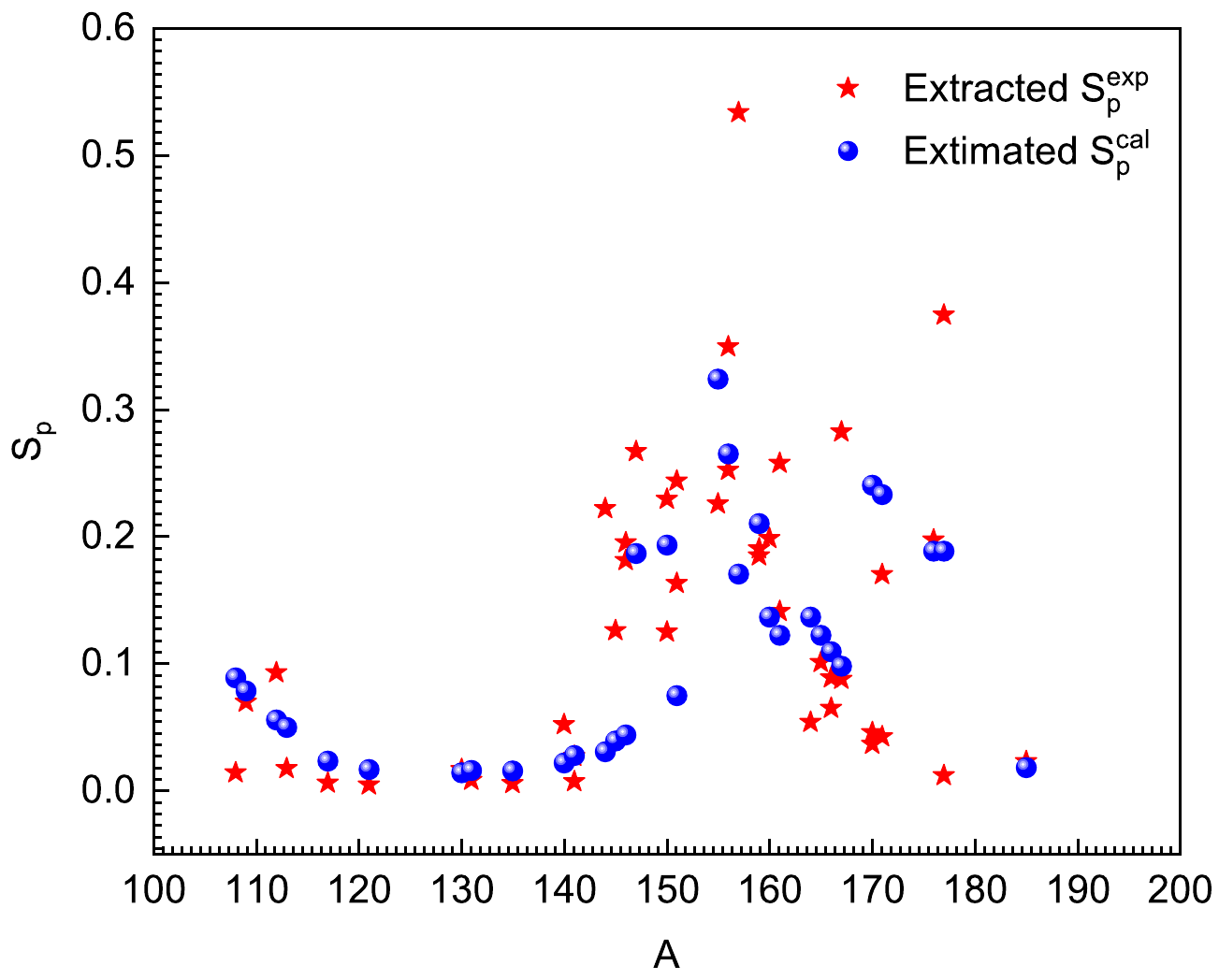}
	\caption{(color online) The experimental spectroscopic factor $S_p^{\rm exp}$ and the calculated one by Eq. \ref{eq16} of deformed proton emitters.}
	\label{fig 2}
\end{figure}

In our previous work \cite{H. F. Gui Commun. Theor. Phys. 2022}, we used the D-TPA to study the effect of daughter deformation on half-life of $\alpha$ decay. Considering proton radioactivity shares the same mechanism as $\alpha$ decay, in this work, we try to extend this model to describe proton radioactivity. Firstly, based on the D-TPA, we calculate the proton radioactivity half-lives with $S_p=1$ for proton emitters in the ground state as well as in the isomeric state. The ratio of calculated proton radioactivity half-life $T_{1/2}^{\rm cal}$ to experimental one $T_{1/2}^{\rm exp}$ is possible to determine the experimental spectroscopic factor ($S_p^{\rm exp})$, which is defined as $S_p^{\rm exp}=T_{1/2}^{\rm cal}/T_{1/2}^{\rm exp}$. A comparison between experimental and theoretical spectroscopic factors has found a good consistency between them\added{,} with the exception of rare cases where the experimental data are below the theoretical predictions \cite{S. Aberg 1997}. These differences in behavior are due to the influence of nuclear structure effects such as deformation and quantum mixing that are not taken into account in simple calculations \cite{P. J. Woods 1997}. In other words, the spectroscopic factor may be dependent on deformation in the microscopic model adopted for calculating the proton radioactivity half-lives. Recently, Delion \emph{et al.} discovered that there is a small residual linear decrease of the spectroscopic factor in logarithmic form  versus the quadrupole deformation parameter of proton emitters($\beta_2^{'}$) \cite{D. S. Delion Phys. Rev. C 2021}. In that regard, we consider establishing the correlation between extracted experimental spectroscopic factors and $\beta_2^{'}$. Adopting the extracted experimental spectroscopic factor in logarithmic form, ${\rm log}_{10}S_p^{\rm exp}$ as a function of $\beta_2^{'}$ is plotted in Fig. \ref{fig 1}. From this figure, it is worth noting there are five points, which clearly deviate from the fitting line, corresponding to particular nuclei $^{108}$I, $^{170}$Au, $^{170}$Au$^m$, $^{171}$Au$^m$ and $^{177}$Tl$^m$. The large discrepancies and small $S_p^{\rm exp}$ values occurs at $Z=79$ and 81, maybe due to the strong proton shell effect. Particularly, the emitted protons are all from $\pi h_{11/2}$ state for $^{171}$Au$^m$ and $^{177}$Tl$^m$, corresponding to the orbit of shell closure in the shell model picture \cite{Qian Y and Ren Eur. Phys. J. A 2016}.  Similarly, the proton number of $^{108}$I is close to the magic number $Z=50$, and the proton shell effect may be a reason for this phenomenon. On the whole, we can clearly see that there is an obvious linear relationship between the experimental spectroscopic factor in logarithmic form ${\rm log_{10}}S_p^{\rm exp}$ and $\beta_2^{'}$. 
Then, a simple formula is put forward for evaluating the spectroscopic factor of proton radioactivity as
\begin{eqnarray}
{\rm log}_{10}S_p=
\left\{
\begin{array}{rr}
a\beta_2^{'}+b & \quad \mbox{for}\ \beta_2^{'}>0,\vspace{0.3em}\\
c\beta_2^{'}+d & \quad \mbox{for}\ \beta_2^{'}<0,\vspace{0.3em} 
\end{array}
\right.
\label{eq16}
\end{eqnarray}
where the parameters $a=-4.353$, $b=-0.398$, $c=1.332$, and $d=-0.479$ are determined by fitting the extracted experimental spectroscopic factor data. In the following, we adopt Eq. \ref{eq16} to calculate the spectroscopic factors, denoted as $S_p^{\rm cal}$. In order to visually exhibit the agreement between the experimental spectroscopic factors $S_p^{\rm exp}$ and calculated ones $S_p^{\rm cal}$, they are plotted as red stars and solid blue circles in Fig. \ref{fig 2}, respectively. From this figure, we can distinctly see that $S_p^{\rm cal}$ can fit $S_p^{\rm exp}$ well, which indicates the simple analytic formula for estimating the spectroscopic factor is credible.
\begin{figure}[htb]\centering
	\includegraphics[width=8.5cm]{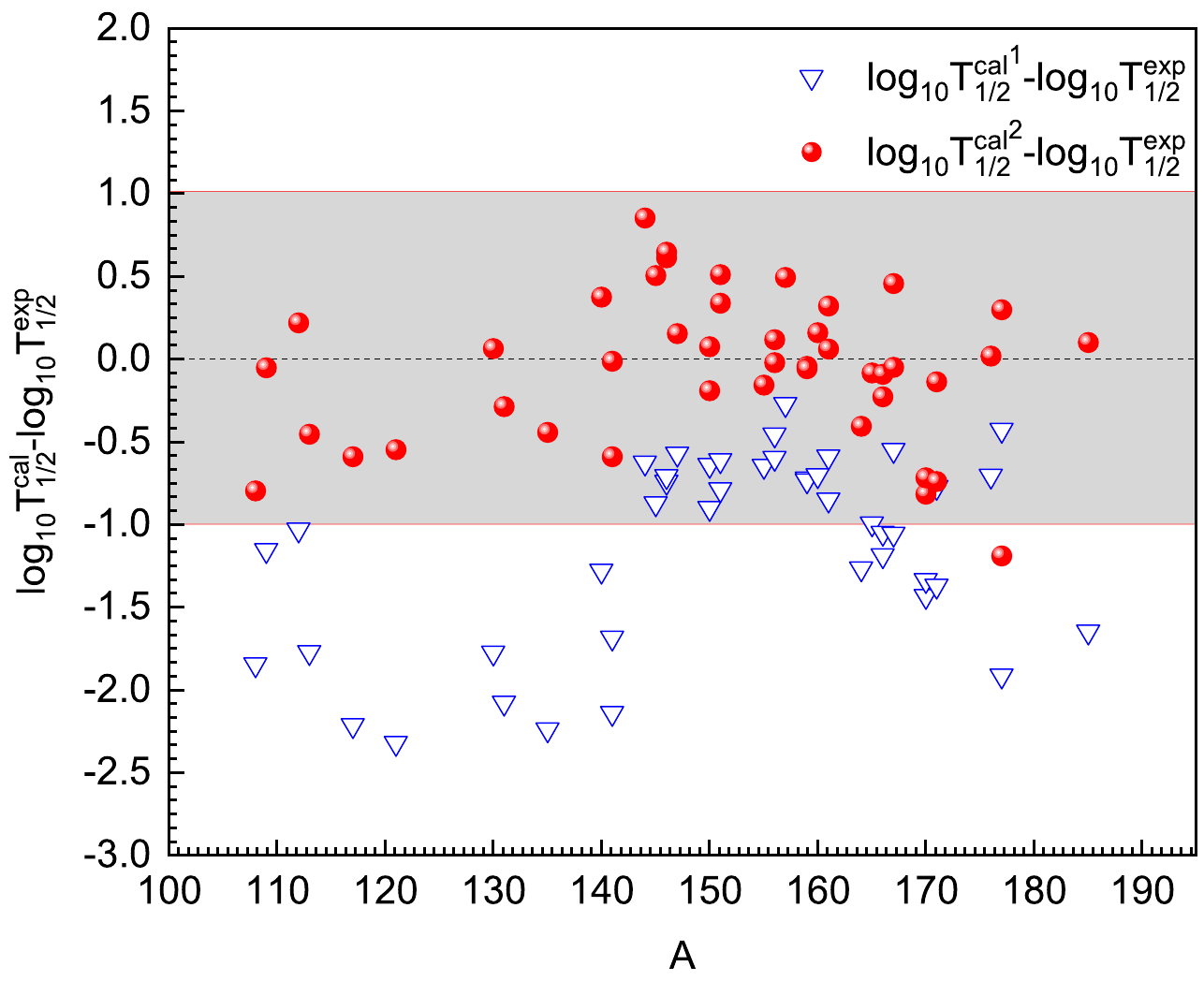}
	\caption{(color online) Deviations between the experimental proton radioactivity half-lives and two calculated ones for deformed nuclei.}
	\label{fig 3}
\end{figure}

Furthermore, based on the D-TPA, we calculate the half-lives of proton radioactivity with the spectroscopic factors taken as 1 and obtained by Eq. \ref{eq16}. For comparison, the universal decay law for proton radioactivity (UDLP) \added{\cite{C. Qi Phys. Rev. C 2012}} and the new Geiger-Nuttall law (NG-N) \cite{Chen J L Eur. Phys. J. A 2019} are also used. All predicted results are listed in Table. \ref{tab1}. In this table, the first five columns represent proton emitters, the released energy of proton radioactivity $Q_p$, the angular momentum $l$ taken away by the emitted proton, extracted experimental spectroscopic factor $S_p^{\rm exp}$, and calculated spectroscopic factor $S_p^{\rm cal}$, respectively.  The last five columns denote the logarithmic form of the experimental proton radioactivity half-lives and theoretical ones calculated using the D-TPA with $S_p$ taken as 1 and $S_p$ obtained by Eq. \ref{eq16}, UDLP and NG-N, which are denoted as Cal$^{1}$, Cal$^{2}$, UDLP and NG-N.
As shown in Table. \ref{tab1}, after considering the spectroscopic factors $S_p^{\rm cal}$ obtained by Eq. \ref{eq16}, calculated proton radioactivity half-lives Cal$^{2}$ can well reproduce experimental data in the region of $10^{-5}$ to $10^2$ s. 
For clarity, the deviations between the experimental proton radioactivity half-lives and the calculated ones ${\rm log}_{10}T_{1/2}^{\rm cal^1}$ and ${\rm log}_{10}T_{1/2}^{\rm cal^2}$ are plotted in Fig. \ref{fig 3}. From this figure, we can clearly see that the ${\rm log}_{10}T_{1/2}^{\rm cal^1}$ are obviously lower than experimental data over the entire region corresponding to the deviations between $-2.5$ and 0.
The reason for large deviations caused by $S_p=1$ maybe that the influence of deformation on spectroscopic factor is not taken into account.
Particularly, the deviations are near zero as well as within $\pm 1$ when the spectroscopic factors obtained by Eq. \ref{eq16} are considered, indicating the ${\rm log}_{10}T_{1/2}^{\rm cal^2}$ can better reproduce experimental data. It explains the correlation between the spectroscopic factor and the deformation effect from a certain point of view and further validates the reliability of our formula.

\begingroup
\renewcommand*{\arraystretch}{1.3}
\setlength\LTleft{0pt}
\setlength\LTright{0pt}
\setlength{\LTcapwidth}{7in}

\begin{longtable*}
	{@{\extracolsep{\fill}} lccccccccc}
	\caption{Comparison of experimental proton radioactivity half-lives with the calculated ones by using different theoretical models and/or formulae. The symbol $m$ denotes the first isomeric state. The experimental proton radioactivity half-lives, spin and parity are taken from Ref. \cite{Kondev F G Chin. Phys. C 2021}. The released energy are given by Eq. \ref{eq7} with the exception of $Q_p$ value for $^{130}$Eu, $^{159}$Re, $^{161}$Re$^{m}$, $^{164}$Ir and $^{177}$Tl$^{m}$ taken from Ref.\cite{Blank B Prog. Part. Nucl. Phys. 2008}.}
	\label{tab1} \\
	\hline 
	\hline 
	{Nucleus}&$Q_p$&$l$&$S_p^{\rm exp}$&$S_p^{\rm cal}$&&&$\rm{log}_{10}{\emph{T}}_{1/2}$(s)&&\\ \cline{6-10} $^{A}Z$&MeV&&&&Exp&Cal$^{1}$&Cal$^{2}$&UDLP\cite{C. Qi Phys. Rev. C 2012}&NG-N\cite{Chen J L Eur. Phys. J. A 2019}\\
	\hline
	\endfirsthead
	\multicolumn{10}{c}%
	{{\tablename\ \thetable{} ($Continued.$)}} \\
	\hline
	\hline 
	{Nucleus}&$Q_p$&$l$&$S_p^{\rm exp}$&$S_p^{\rm cal}$&&&$\rm{log}_{10}{\emph{T}}_{1/2}$(s)&&\\ \cline{6-10} $^{A}Z$&MeV&&&&Exp&Cal$^{1}$&Cal$^{2}$&UDLP\cite{C. Qi Phys. Rev. C 2012}
	&NG-N\cite{Chen J L Eur. Phys. J. A 2019}\\
	\hline
	\endhead
	\hline  \\
	\endfoot
	\hline \hline
	\endlastfoot
$	^	{	108	}	$I	&	0.610 	&	2	&	0.014 	&	0.089 	&	0.723 	&	--1.124 	&	--0.073 	&	--0.019 	&	0.829 	\\
$	^	{	109	}	$I	&	0.829 	&	2	&	0.070 	&	0.079 	&	--4.032 	&	--5.187 	&	--4.084 	&	--3.671 	&	--3.157 	\\
$	^	{	112	}	$Cs	&	0.820 	&	2	&	0.093 	&	0.056 	&	--3.310 	&	--4.341 	&	--3.090 	&	--2.923 	&	--2.362 	\\
$	^	{	113	}	$Cs	&	0.981 	&	2	&	0.018 	&	0.050 	&	--4.771 	&	--6.542 	&	--5.224 	&	--4.899 	&	--4.490 	\\
$	^	{	117	}	$La	&	0.831 	&	2	&	0.006 	&	0.024 	&	--1.664 	&	--3.878 	&	--2.252 	&	--2.459 	&	--1.857 	\\
$	^	{	121	}	$Pr	&	0.901 	&	2	&	0.005 	&	0.017 	&	--1.921 	&	--4.241 	&	--2.468 	&	--2.811 	&	--2.237 	\\
$	^	{	130	}	$Eu	&	1.039 	&	2	&	0.021 	&	0.014 	&	--3.000 	&	--4.776 	&	--2.936 	&	--3.357 	&	-2.789 	\\
$	^	{	131	}	$Eu	&	0.963 	&	2	&	0.008 	&	0.016 	&	--1.699 	&	--3.777 	&	--1.986 	&	--2.458 	&	--1.840 	\\
$	^	{	135	}	$Tb	&	1.203 	&	3	&	0.006 	&	0.016 	&	--2.996 	&	--5.235 	&	--3.437 	&	--3.806 	&	--3.212 	\\
$	^	{	140	}	$Ho	&	1.104 	&	3	&	0.052 	&	0.022 	&	--2.222 	&	--3.502 	&	--1.846 	&	--2.317 	&	--1.600 	\\
$	^	{	141	}	$Ho	&	1.194 	&	3	&	0.007 	&	0.028 	&	--2.387 	&	--4.528 	&	--2.977 	&	--3.257 	&	--2.583 	\\
$	^	{	141	}	$Ho$^m$	&	1.264 	&	0	&	0.027 	&	0.028 	&	--5.137 	&	--6.822 	&	--5.147 	&	--5.331 	&	--5.524 	\\
$	^	{	144	}	$Tm	&	1.724 	&	5	&	0.222 	&	0.031 	&	--5.569 	&	--6.194 	&	--4.714 	&	--4.687 	&	--4.991 	\\
$	^	{	145	}	$Tm	&	1.754 	&	5	&	0.126 	&	0.039 	&	--5.499 	&	--6.371 	&	--4.994 	&	--4.871 	&	--5.182 	\\
$	^	{	146	}	$Tm	&	0.904 	&	0	&	0.181 	&	0.044 	&	--0.810 	&	--1.551 	&	--0.196 	&	--0.610 	&	--0.968 	\\
$	^	{	146	}	$Tm$^{m}$	&	1.214 	&	5	&	0.195 	&	0.044 	&	--1.137 	&	--1.846 	&	--0.490 	&	--0.896 	&	--0.737 	\\
$	^	{	147	}	$Tm$^{m}$	&	1.133 	&	2	&	0.267 	&	0.187 	&	--3.444 	&	--4.017 	&	--3.289 	&	--2.859 	&	--2.183 	\\
$	^	{	150	}	$Lu	&	1.285 	&	5	&	0.230 	&	0.193 	&	--1.347 	&	--1.986 	&	--1.272 	&	--1.132 	&	--0.965 	\\
$	^	{	150	}	$Lu$^{m}$	&	1.305 	&	2	&	0.125 	&	0.193 	&	--4.398 	&	--5.298 	&	--4.588 	&	--4.050 	&	--3.381 	\\
$	^	{	151	}	$Lu	&	1.255 	&	5	&	0.163 	&	0.075 	&	--0.896 	&	--1.682 	&	--0.558 	&	--0.862 	&	--0.653 	\\
$	^	{	151	}	$Lu$^{m}$	&	1.315 	&	2	&	0.244 	&	0.075 	&	--4.796 	&	--5.408 	&	--4.284 	&	--4.150 	&	--3.471 	\\
$	^	{	155	}	$Ta	&	1.466 	&	5	&	0.226 	&	0.324 	&	--2.495 	&	--3.141 	&	--2.652 	&	--2.269 	&	--2.161 	\\
$	^	{	156	}	$Ta	&	1.036 	&	2	&	0.252 	&	0.265 	&	--0.826 	&	--1.424 	&	--0.848 	&	--0.624 	&	0.102 	\\
$	^	{	156	}	$Ta$^{m}$	&	1.126 	&	5	&	0.349 	&	0.265 	&	0.933 	&	0.476 	&	1.053 	&	0.947 	&	1.371 	\\
$	^	{	157	}	$Ta	&	0.946 	&	0	&	0.534 	&	0.171 	&	--0.527 	&	--0.800 	&	--0.032 	&	--0.038 	&	--0.363 	\\
$	^	{	159	}	$Re	&	1.816 	&	5	&	0.190 	&	0.211 	&	--4.678 	&	--5.400 	&	--4.721 	&	--4.270 	&	--4.285 	\\
$	^	{	159	}	$Re$^{m}$	&	1.816 	&	5	&	0.185 	&	0.211 	&	--4.665 	&	--5.398 	&	--4.721 	&	--4.269 	&	--4.283 	\\
$	^	{	160	}	$Re	&	1.276 	&	2	&	0.199 	&	0.137 	&	--3.163 	&	--3.865 	&	--3.001 	&	--2.841 	&	--2.101 	\\
$	^	{	161	}	$Re	&	1.216 	&	0	&	0.258 	&	0.123 	&	--3.357 	&	--3.945 	&	--3.034 	&	--2.895 	&	--3.018 	\\
$	^	{	161	}	$Re$^{m}$	&	1.338 	&	5	&	0.141 	&	0.123 	&	--0.678 	&	--1.528 	&	--0.616 	&	--0.806 	&	--0.501 	\\
$	^	{	164	}	$Ir	&	1.844 	&	5	&	0.054	&	0.137 	&	--3.947 	&	--5.213 	&	--4.351 	&	--4.114 	&	--4.039 	\\
$	^	{	165	}	$Ir$^{m}$	&	1.727 	&	5	&	0.101	&	0.123 	&	--3.433 	&	--4.427 	&	--3.515 	&	--3.408 	&	--3.266 	\\
$	^	{	166	}	$Ir	&	1.167 	&	2	&	0.065 	&	0.110 	&	--0.824 	&	--2.009 	&	--1.050 	&	--1.188 	&	--0.426 	\\
$	^	{	166	}	$Ir$^{m}$	&	1.347 	&	5	&	0.089 	&	0.110 	&	--0.076 	&	--1.126 	&	--0.166 	&	--0.475 	&	--0.100 	\\
$	^	{	167	}	$Ir	&	1.087 	&	0	&	0.283 	&	0.098 	&	--1.120 	&	--1.668 	&	--0.661 	&	--0.865 	&	--1.076 	\\
$	^	{	167	}	$Ir$^{m}$	&	1.262 	&	5	&	0.088 	&	0.098 	&	0.842 	&	--0.214	&	0.794 	&	0.348 	&	0.798 	\\
$	^	{	170	}	$Au	&	1.487 	&	2	&	0.037 	&	0.241 	&	--3.487 	&	--4.920 	&	--4.302 	&	--3.845 	&	--3.023 	\\
$	^	{	170	}	$Au$^{m}$	&	1.767 	&	5	&	0.046 	&	0.241 	&	--2.971 	&	--4.307 	&	--3.688 	&	--3.333 	&	--3.118 	\\
$	^	{	171	}	$Au	&	1.464 	&	0	&	0.170 	&	0.233 	&	--4.652 	&	--5.425 	&	--4.788 	&	--4.298 	&	--4.228 	\\
$	^	{	171	}	$Au$^{m}$	&	1.718 	&	5	&	0.043 	&	0.233 	&	--2.587 	&   --3.957 	&	--3.324 	&	--3.026 	&	--2.777 	\\
$	^	{	176	}	$Tl	&	1.278 	&	0	&	0.198 	&	0.189 	&	--2.208 	&	--2.912 	&	--2.187 	&	--2.059 	&	--2.113 	\\
$	^	{	177	}	$Tl	&	1.173 	&	0	&	0.375 	&	0.189 	&	--1.174 	&	--1.600 	&	--0.876 	&	--0.875 	&	--1.015 	\\
$	^	{	177	}	$Tl$^{m}$	&	1.967 	&	5	&	0.012 	&	0.189 	&	--3.346 	&	--5.260 	&	--4.535 	&	--4.227 	&	--3.972 	\\
$	^	{	185	}	$Bi$^{m}$	&	1.625 	&	0	&	0.023 	&	0.018 	&	--4.191 	&	--5.838 	&	--4.090 	&	--4.759 	&	--4.511 	\\
\end{longtable*}
\endgroup

\begin{table}[!htb] 
	\renewcommand\arraystretch{1.5}
	\caption{Standard deviations $\sigma$ between experimental half-lives and the calculated ones using different theoretical models and/or formulae.} 
	\label{tab2} 
	\centering
	\begin{tabular*}{8cm} {@{\extracolsep{\fill}} ccccc}
		\hline
		\hline
		Type & Cal$^{1}$ & Cal$^{2}$ & UDLP & NG-N 
		\\  
		\hline 
		$\sigma$  & 1.253 & 0.443 &0.478
		& 0.516 \\	 
		\hline
		\hline
	\end{tabular*}  
\end{table}
\begin{figure}[htb]\centering
	\includegraphics[width=8.5cm]{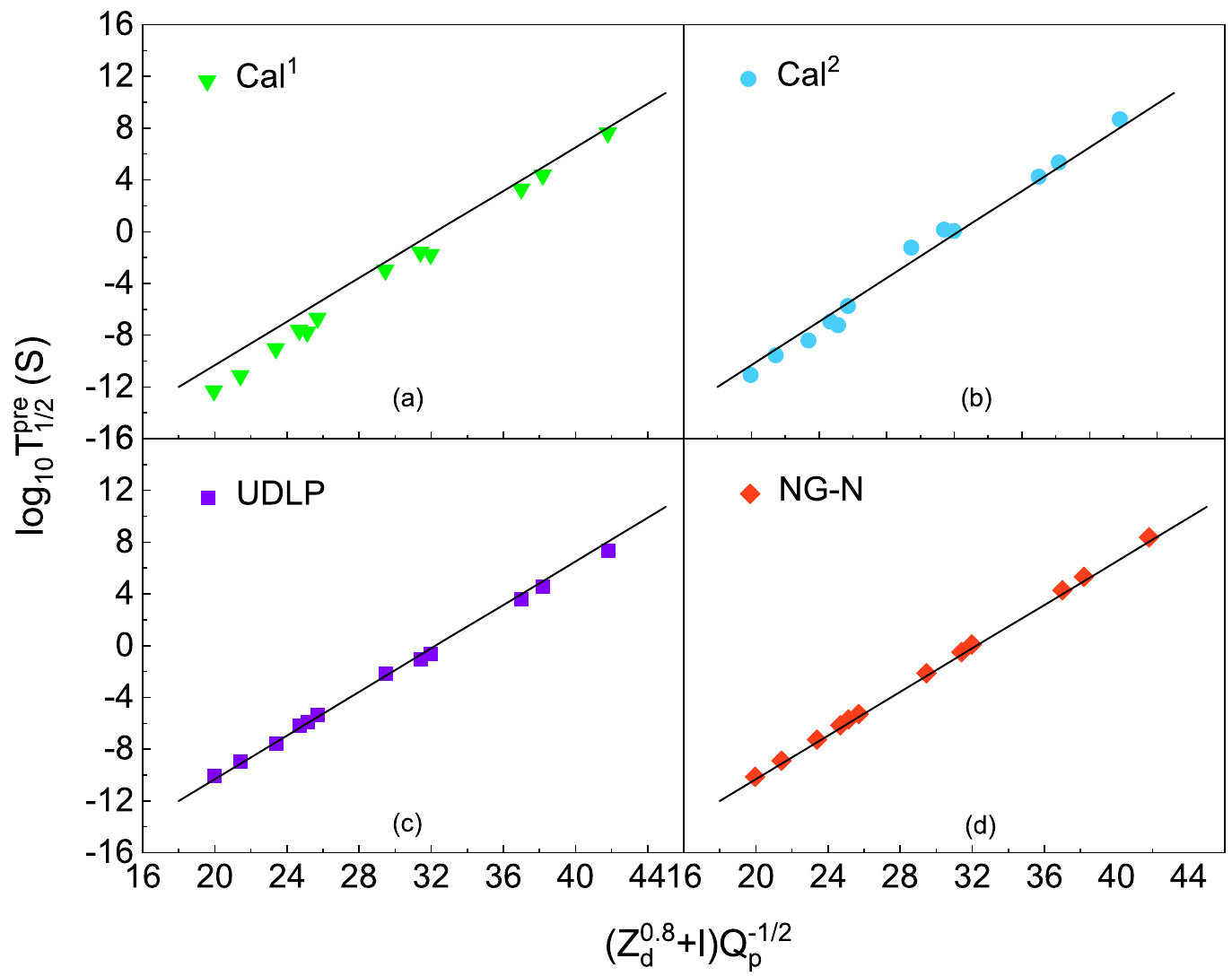}
	\caption{(color online) Relationship  between  the  predicted results of these models and/or formulae listed in Table. \ref{tab3} and coulomb parameters considering the effect of the orbital angular momentum, i.e., NG-N \cite{Chen J L Eur. Phys. J. A 2019}.}
	\label{fig 4}
\end{figure}
In order to gain a global insight into the agreement between experimental data and the calculated ones, the root mean square deviation $\sigma$ is used to quantify the calculated capabilities of the above models and/or formulae. It can be defined as
\begin{eqnarray}
\sigma=\sqrt{\frac{1}{N}\sum_{i=1}^N{({\rm log}_{10}{T_{1/2}^{{\rm cal}.i}}-{\rm log}_{10}{T_{1/2}^{{\rm exp}.i}})^2}},
\label{eq17}
\end{eqnarray}
where ${\rm log}_{10}{T_{1/2}^{{\rm cal}.i}}$ and ${\rm log}_{10}{T_{1/2}^{{\rm exp}.i}}$ denote the experimental proton radioactivity half-life and the calculated one in logarithmic form for the $i$-th nucleus, respectively. For comparison, all the specific results of $\sigma$ are listed in Table. \ref{tab2}, which show that $\sigma_{{\rm Cal}^{1}}=1.253$, $\sigma_{{\rm Cal}^{2}}=0.443$, $\sigma_{\rm UDLP}=0.478$ and $\sigma_{\rm NG-N}=0.516$. These results indicate that there is indeed some improvement after considering the spectroscopic factor. The spectroscopic factor of proton radioactivity can be simply described by a formula of $\beta_2^{'}$. Besides,  we can see that the $\sigma_{{\rm Cal}^{2}}=0.443$ is smaller than the ones calculated by the UDLP and NG-N, which are reduced by $7.32\%$ and $14.15\%$.

Given the good agreement between the calculated results with the experimental data, based the D-TPA, we use the present formula to estimate the spectroscopic factors and predict half-lives of 12 possible proton emitters whose radioactivity is energetically allowed or observed but not quantified in NUBASE2020 \cite{Kondev F G Chin. Phys. C 2021}. For comparison, the UDLP and NG-N are also used. The detailed predictions are listed in the Table. \ref{tab3}. In this table, the first four columns represent the possible proton radioactivity candidates, released energy $Q_p$, orbital angular momentum $l$, and calculated spectroscopic factors $S_p^{\rm cal}$, respectively. The next two columns represent the calculated proton radioactivity half-lives obtained by our model with $S_p=1$ and after considering the spectroscopic factor in logarithmic form, denoted as Cal$^1$ and Cal$^2$, respectively. The last two columns are the predicted proton radioactivity half-lives using UDLP and NG-N. 
Note from this table that the predicted proton radioactivity half-lives using our model are reasonably consistent with the ones using the other two formulae. 
Moreover, to further compare the evaluation capabilities of the above approaches, the relationship between the  predicted results of these four models and/or formulae and coulomb parameters considering the effect of the orbital angular momentum, i.e., NG-N, is plotted in Fig. \ref{fig 4}. The results show that all predictions are basically consistent with the NG-N, which means our predictions of the proton radioactivity half-lives are valid and reliable. These predictions may provide valuable information for searching the new nuclide with the proton radioactivity in the future.
\begingroup
\renewcommand*{\arraystretch}{1.3}
\setlength\LTleft{0pt}
\setlength\LTright{0pt}
\setlength{\LTcapwidth}{7in}
\begin{longtable*}
	{@{\extracolsep{\fill}} lccccccc}
	\caption{Comparison of the predicted half-lives for possible radioactivity candidates whose proton radioactivity is energetically allowed or observed but not yet quantified in NUBASE2020. The symbol $m$ denotes the first isomeric state. The released energy of proton radioactivity are given by Eq. \ref{eq7}.}
	\label{tab3} \\
	\hline 
	\hline 
	{Nucleus}&$Q_p$&$l$&$S_p^{\rm cal}$&&$\rm{log}_{10}{\emph{T}}_{1/2}$(s)&&\\		\cline{5-8}		$^{A}Z$&MeV&&&Cal$^1$&Cal$^2$&UDLP\cite{C. Qi Phys. Rev. C 2012}&NG-N\cite{Chen J L Eur. Phys. J. A 2019}	\\
	\hline
	\endfirsthead
	\multicolumn{8}{c}%
	{{\tablename\ \thetable{} ($Continued.$)}} \\
	\hline
	\hline 
	{Nucleus}&$Q_p$&$l$&$S_p$&&$\rm{log}_{10}{\emph{T}}_{1/2}$(s)&&\\		\cline{5-8}		$^{A}Z$&MeV&&&Cal$^1$&Cal$^2$&UDLP\cite{C. Qi Phys. Rev. C 2012}&NG-N\cite{Chen J L Eur. Phys. J. A 2019}	\\
	\hline
	\endhead
	\hline  \\
	\endfoot
	\hline \hline
	\endlastfoot	
	$	^	{103}	$Sb	&	0.979 	&	2	&0.324&	--7.732 	&	--7.243 	&	--5.948 	&	--5.698 	\\
	$	^	{111}	$Cs	&	1.740 	&	2	&0.063&	--12.292 	&	--11.088 	&	--10.094 	&	--10.145 	\\
	$	^	{116}	$La	&	1.591 	&	2	&0.026&	--11.117 	&	--9.539 	&	--9.000 	&	--8.887 	\\
	$	^	{127}	$Pm	&	0.792 	&	2	&0.015&	--1.761 	&	0.069 	&	--0.620 	&	0.094 	\\
	$	^	{159}	$Re	&	1.606 	&	0	&0.211&	--7.614 	&	--6.937 	&	--6.227 	&	--6.156 	\\
	$	^	{162}	$Re	&	0.760 	&	2	&0.110&	4.388 	&	5.348 	&	4.597 	&	5.322 	\\
	$	^	{165}	$Ir	&	1.547 	&	0	&0.123&	--6.662 	&	--5.750 	&	--5.387 	&	--5.303 	\\
	$	^	{169}	$Ir$^{\rm m}$	&	0.782 	&	5	&0.088&	7.638 	&	8.692 	&	7.362 	&	8.363 	\\
	$	^	{169}	$Au	&	1.947 	&	0	&0.241&	--9.027 	&	--8.408 	&	--7.572 	&	--7.276 	\\
	$	^	{172}	$Au	&	0.877 	&	2	&0.123&	3.319 	& 4.231 	&	3.578 	&	4.284 	\\
	$	^	{184}	$Bi	&	1.330 	&	0	&0.018&	--2.949 	&--1.215 	&	--2.144 	&	--2.123 	\\
	$	^	{185}	$Bi	&	1.541 	&	5	&0.018&	--1.568 	&	0.167 	&	--1.030 	&	--0.508 	\\	
\end{longtable*}
\endgroup

\section{Summary}
\label{section 4}
In summary, for evaluating the spectroscopic factor of proton radioactivity, we propose a simple analytic expression for describing the relationship between the spectroscopic factor extracted from the experimental  proton radioactivity half-life and the quantity $\beta_2^{'}$. 
The present formula for the spectroscopic factor can be successfully applied to estimate the spectroscopic factors as well as calculate half-lives. As an application, we extend this model to predict proton radioactivity half-lives for 12 possible candidates. The predicted results are in great agreement with other ones obtained by UDLP and NG-N. This work may provide useful and reliable information for experimental and theoretical research in the future.

\begin{acknowledgments}
This work is supported in part by the National Natural Science Foundation of China (Grant No.12175100 and No.11975132), the construct program of the key discipline in hunan province, the Research Foundation of Education Bureau of Hunan Province, China (Grant No.18A237), the Shandong Province Natural Science Foundation, China (Grant No.ZR2022JQ04), the Opening Project of Cooperative Innovation Center for Nuclear Fuel Cycle Technology and Equipment, University of South China (Grant No.2019KFZ10), the Innovation Group of Nuclear and Particle Physics in USC, Hunan Provincial Innovation Foundation for Postgraduate (Grant No.CX20210942).
\end{acknowledgments}


\begin{thebibliography}{91}%
\makeatletter
\providecommand \@ifxundefined [1]{%
 \@ifx{#1\undefined}
}%
\providecommand \@ifnum [1]{%
 \ifnum #1\expandafter \@firstoftwo
 \else \expandafter \@secondoftwo
 \fi
}%
\providecommand \@ifx [1]{%
 \ifx #1\expandafter \@firstoftwo
 \else \expandafter \@secondoftwo
 \fi
}%
\providecommand \natexlab [1]{#1}%
\providecommand \enquote  [1]{``#1''}%
\providecommand \bibnamefont  [1]{#1}%
\providecommand \bibfnamefont [1]{#1}%
\providecommand \citenamefont [1]{#1}%
\providecommand \href@noop [0]{\@secondoftwo}%
\providecommand \href [0]{\begingroup \@sanitize@url \@href}%
\providecommand \@href[1]{\@@startlink{#1}\@@href}%
\providecommand \@@href[1]{\endgroup#1\@@endlink}%
\providecommand \@sanitize@url [0]{\catcode `\\12\catcode `\$12\catcode
  `\&12\catcode `\#12\catcode `\^12\catcode `\_12\catcode `\%12\relax}%
\providecommand \@@startlink[1]{}%
\providecommand \@@endlink[0]{}%
\providecommand \url  [0]{\begingroup\@sanitize@url \@url }%
\providecommand \@url [1]{\endgroup\@href {#1}{\urlprefix }}%
\providecommand \urlprefix  [0]{URL }%
\providecommand \Eprint [0]{\href }%
\providecommand \doibase [0]{http://dx.doi.org/}%
\providecommand \selectlanguage [0]{\@gobble}%
\providecommand \bibinfo  [0]{\@secondoftwo}%
\providecommand \bibfield  [0]{\@secondoftwo}%
\providecommand \translation [1]{[#1]}%
\providecommand \BibitemOpen [0]{}%
\providecommand \bibitemStop [0]{}%
\providecommand \bibitemNoStop [0]{.\EOS\space}%
\providecommand \EOS [0]{\spacefactor3000\relax}%
\providecommand \BibitemShut  [1]{\csname bibitem#1\endcsname}%
\let\auto@bib@innerbib\@empty
\bibitem{C. Xu Phys. Rev. C 2016} C. Xu and Z. Z. Ren, \href{https://doi.org/10.1103/PhysRevC.73.041301}{Phys. Rev. C {\bf 73}, 041301(R) (2016)}.
 
\bibitem{Delion D S Phy. Rev. Lett. 2006} D. S. Delion, R. J. Liotta, and R. Wyss, \href{https://doi.org/10.1103/PhysRevLett.96.072501}{Phy. Rev. Lett. {\bf 96}, 072501 (2006)}.

\bibitem{Blank B Prog. Part. Nucl. Phys. 2008} B. Blank and M. J. G. Borge, \href{https://doi.org/10.1016/j.ppnp.2007.12.001}{Prog. Part. Nucl. Phys. {\bf 60}, 403 (2008)}.

\bibitem{Zhang H F J. Phys. G: Nucl. Part. Phys. 2010} H. F. Zhang, Y. J. Wang, J. M. Dong, J. Q. Li, and W. Scheid, \href{https://doi.org/10.1088/0954-3899/37/8/085107}{J. Phys. G: Nucl. Part. Phys. {\bf 37}, 085107 (2010)}.

\bibitem{Chen J L J. Phys. G: Nucl. Part. Phys. 2019} J. L. Chen, X. H. Li, J. H. Cheng, J. G. Deng, and X. J. Wu, \href{https://doi.org/10.1088/1361-6471/ab1a56}{J. Phys. G: Nucl. Part. Phys. {\bf 46}, 065107 (2019)}.

\bibitem{Delion D S Phys. Rev. C 2009} D. S. Delion,  \href{https://doi.org/10.1103/PhysRevC.80.024310}{Phys. Rev. C {\bf 80}, 024310 (2009)}.

\bibitem{Karny M Phys. Lett. B 2008} M. Karny, K. P. Rykaczewski, R. K. Grzywacz, J.C. Batchelder, C.R. Bingham, C. Goodin, C.J. Gross, J.H. Hamilton, A. Korgul, W. Kr\'{o}las, S.N. Liddick, K. Li, K.H. Maier, C. Mazzocchi, A. Piechaczek, K. Rykaczewski, D. Schapira, D. Simpson, M.N. Tantawy, J.A. Winger, and M.V. Stoitsov, \href{https://doi.org/10.1016/j.physletb.2008.04.056}{Phys. Lett. B {\bf 664}, 52 (2008)}.

\bibitem{Zhang Z X Chin. Phys. C 2018} Z. X. Zhang and J. M. Dong, \href{https://doi.org/10.1088/1674-1137/42/1/014104}{Chin. Phys. C {\bf 42}, 014104 (2018)}.

\bibitem{Basu D N Phys. Rev. C 2005} D. N. Basu, P. R. Chowdhury, and C. Samanta, \href{https://doi.org/10.1103/PhysRevC.72.051601}{Phys. Rev. C {\bf 72}, 051601 (2005)}.

\bibitem{Jackson K Phys. Lett. B 1970} K. P. Jackson, C. U. Cardinal, H. C. Evans, N. A. Jelley, and J. Cerny, \href{https://doi.org/10.1016/0370-2693(70)90269-8}{Phys. Lett. B {\bf 33}, 281 (1970)}.

\bibitem{Cerny J Phys. Lett. B 1970} J. Cerny, J. Esterl, R. Gough, and R. Sextro, \href{https://doi.org/10.1016/0370-2693(70)90270-4}{Phys. Lett. B {\bf 33}, 284 (1970)}.	

\bibitem{Hofmann S Z. Phys. A 1982} S. Hofmann, W. Reisdorf, G. M\"{u}nzenberg, F. P. He\ss berger, J. R. H. Schneider, and P. Armbruster, \href{https://doi.org/10.1007/BF01415018}{Z. Phys. A {\bf 305}, 111 (1982)}.

\bibitem{O. Klepper Z. Phys. A: At. Nucl. 1982} O. Klepper, T. Batsch, S. Hofmann, R. Kirchner, W. Kurcewicz, W. Reisdorf, E. Roeckl, D. Schardt, and G. Nyman, \href{https://doi.org/10.1007/BF01415019}{Z. Phys. A: At. Nucl. {\bf 305}, 125 (1982)}.

\bibitem{T. Faestermann Phys. Lett. B 1984} T. Faestermann, A. Gillitzer, K. Hartel, P. Kienle, and E. Nolte, \href{https://doi.org/10.1016/0370-2693(84)91098-0}{Phys. Lett. B {\bf 137}, 23 (1984)}.

\bibitem{S. Hofmann 1984} S. Hofmann, in \textit{Proceedings of the 7th International Conference on Atomic Masses Fundamental Constants,} edited by O.Klepper (Darmstadt, Germany, 1984), Vol. 26, p. 184.

\bibitem{C. N. Davids Phys. Rev. Lett. 1998} C. N. Davids, P. J. Woods, D. Seweryniak, A. A. Sonzogni, J. C. Batchelder, C. R. Bingham, T. Davinson, D. J. Henderson, R. J. Irvine, G. L. Poli, J. Uusitalo, and W. B. Walters, \href{https://doi.org/10.1103/PhysRevLett.80.1849}{Phys. Rev. Lett. {\bf 80}, 1849 (1998)}.

\bibitem{Santhosh K P Pramana J. Phys.} K. P. Santhosh and I. Sukumaran, \href{https://doi.org/10.1007/s12043-018-1672-4}{Pramana J. Phys. {\bf 92}, 6 (2019)}.

\bibitem{Zhang H F Sci. China. Ser. G:Phy. Mech. Astron.} H. F. Zhang, Y. Z. Wang, J. M. Dong, and J. Q. Li, \href{https://doi.org/10.1007/s11433-009-0204-0}{Sci. China Ser. G-Phys. Mech. Astron. {\bf 52}, 1536-1541 (2009)}.

\bibitem{Y. Z. Wang Commun. Theor. Phys. 2021} Y. Z. Wang, J. P. Cui, Y. H. Gao, and J. Z. Gu, \href{https://doi.org/10.1088/1572-9494/abfa00}{Commun. Theor. Phys. {\bf 73(7)}, 075301 (2021)}.

\bibitem{Qian Y and Ren Eur. Phys. J. A 2016} Y. B. Qian and Z. Z. Ren, \href{https://doi.org/10.1140/epja/i2016-16068-3}{Eur. Phys. J. A {\bf 52}, 68 (2016)}.

\bibitem{L.S. Ferreira PhysRevC.65.024323}  L.S. Ferreira, E. Maglione, and D. E. P. Fernandes, \href{https://doi.org/10.1103/PhysRevC.65.024323}{Phys. Rev. C {\bf 65}, 024323 (2002)}.

\bibitem{P. J. Woods 1997} P. J. Woods and C. N. Davids,  \href{https://doi.org/10.1146/annurev.nucl.47.1.541}{Annu. Rev. Nucl. Part. Sci. {\bf 47(1)}, 541-590 (1997)}.

\bibitem{J. M. Dong Phys. Rev. C 2009} J. M. Dong, H. F. Zhang, and G. Royer, \href{https://doi.org/10.1103/PhysRevC.79.054330}{Phys. Rev. C {\bf 79}, 054330 (2009)}.

\bibitem{D. Delion Phys. Rep. 2006} D. S. Delion, R. J. Liotta, and R. Wyss, \href{https://doi.org/10.1016/j.physrep.2005.11.001}{Phys. Rep. {\bf 80}, 424 (2006)}.

\bibitem{A.Soylu Chin. Phys. C 2021} A. Soylu, F. Koyuncu, G. Gangopadhyay, V. Dehghani, and S. A. Alavi, \href{https://doi.org/10.1088/1674-1137/abe03f}{Chin. Phys. C {\bf 45}, 044108 (2021)}.

\bibitem{Y. Lim Phys. Rev. C 2016} Y. Lim, X. Xia, and Y. Kim, \href{https://doi.org/10.1103/PhysRevC.93.014314}{Phys. Rev. C {\bf 93}, 014314 (2016)}.

\bibitem{Q. Zhao Phys. Rev. C 2014} Q. Zhao, J. M. Dong, J. L. Song, and W.H. Long, \href{https://doi.org/10.1103/PhysRevC.90.054326}{Phys. Rev. C {\bf 90}, 054326 (2014)}.

\bibitem{Chen J L et al Eur. Phys. J. A 2021} J. L. Chen, X. H. Li, X. J. Wu, P. C. Chu, and B. He, \href{https://doi.org/10.1140/epja/s10050-021-00618-1}{Eur. Phys. J. A {\bf 57}, 305 (2021)}.

\bibitem{D. S. Delion Phys. Rev. C 2021} D. S. Delion and A. Dumitrescu, \href{https://doi.org/10.1103/PhysRevC.103.054325}{Phys. Rev. C {\bf 103}, 054325 (2021)}.

\bibitem{J. H Cheng Phys. Rev. C 2022} J. H. Cheng, Y. Li, and T. P. Yu, \href{https://doi.org/10.1103/PhysRevC.105.024312}{Phys. Rev. C {\bf 105}, 024312 (2022)}.

\bibitem{H. F. Gui Commun. Theor. Phys. 2022} H. F. Gui, H. M. Liu, X. J. Wu, P. C. Chu, B. He, and X. H. Li, \href{https://doi.org/10.1088/1572-9494/ac6576}{Commun. Theor. Phys. {\bf 74}, 055301 (2022)}.

\bibitem{Kondev F G Chin. Phys. C 2021} F. G. Kondev, M. Wang, W. J. Huang, S. Naimi, and G. Audi, \href{https://doi.org/10.1088/1674-1137/abddae}{Chin. Phys. C {\bf 45}, 030001 (2021)}.

\bibitem{Denisov V Y Phys. Rev. C 2005} V. Y. Denisov and H. Ikezoe, \href{https://doi.org/10.1103/PhysRevC.72.064613}{Phys. Rev. C \textbf{72}, 064613 (2005).}

\bibitem{K. N. Huang 1976} K. N. Huang, M. Aoyagi, M. H. Chen, B. Crasemann, and H. Mark, \href{https://doi.org/10.1016/0092-640X(76)90027-9}{At. Data Nucl. Data Tables {\bf 18}, 243 (1976)}.

\bibitem{B. Buck Phys. Rev. C 1992} B. Buck, A. C. Merchant, and S. M. Perez, \href{https://doi.org/10.1103/PhysRevC.45.2247}{Phys. Rev. C {\bf 45}, 2247 (1992)}.

\bibitem{M. Ismail  Nucl. Phys. A 2017} M. Ismail, W. M. Seif, A. Adel, and A. Abdurrahman, \href{https://doi.org/10.1016/j.nuclphysa.2016.11.010}{Nucl. Phys. A {\bf 958}, 202-10 (2017)}.

\bibitem{A. J. Sierk 2016} P. M$\ddot o$ller, A. J. Sierk, T. Ichikawa, and H. Sagawa, \href{https://doi.org/10.1016/j.adt.2015.10.002}{At. Data Nucl. Data {\bf 109}, 1-204 (2016)}.

\bibitem{N. Wang and W. Scheid Phys. Rev. C 2008} N. Wang and W. Scheid,  \href{https://doi.org/10.1103/PhysRevC.78.014607}{Phys. Rev. C {\bf 78}, 014607 (2008)}.

\bibitem{N. Takigawa Phys. Rev. C 2000} N. Takigawa, T. Rumin, and N. Ihara, \href{https://doi.org/10.1103/PhysRevC.61.044607}{Phys. Rev. C {\bf 61}, 044607 (2000)}.

\bibitem{M. Ismail Phys. Lett. B 2003} M. Ismail, W. M. Seif, and H. El-Gebaly, \href{https://doi.org/10.1016/S0370-2693(03)00600-2}{Phys. Lett. B {\bf 563}, 53-6 (2003)}.

\bibitem{G. L. Zhang Chin. Phys. Lett. 2008} G. L. Zhang, X. Y. Le, and Z. H. Liu, \href{https://doi.org/10.1088/0256-307X/25/4/023}{Chin. Phys. Lett. {\bf 25}, 1247 (2008)}.

\bibitem{Morehead J J J. Math. Phys. 1995} J. J. Morehead, \href{https://doi.org/10.1063/1.531270}{J. Math. Phys. {\bf 36}, 5431 (1995).}

\bibitem{S. Aberg 1997} S. $\r{A}$berg, P. B. Semmes, and W. Nazarewicz, \href{https://doi.org/10.1103/PhysRevC.56.1762}{Phys. Rev. C {\bf 56}, 1762 (1997)}.

\bibitem{C. Qi Phys. Rev. C 2012} C. Qi, D.S. Delion, R.J. Liotta, and R. Wyss, \href{https://doi.org/10.1103/PhysRevC.85.011303}{Phys. Rev. C {\bf 85}, 011303 (2012)}.

\bibitem{Chen J L Eur. Phys. J. A 2019} J. L. Chen, J. Y. Xu, J. G. Deng, X. H. Li, B. He, and P. C. Chu, \href{https://doi.org/10.1140/epja/i2019-12927-7}{Eur. Phys. J. A {\bf 55}, 214 (2019)}.
\end{thebibliography}

%

\end{document}